\documentclass[preprint,10pt]{elsarticle}
\usepackage[a4paper,left=2.5cm,right=2.5cm,top=2cm,bottom=2cm]{geometry}
\usepackage{amssymb,graphicx,nicefrac,amsmath,tabularx,booktabs,longtable,bm,multirow,multicol,array,graphicx,subcaption,algorithm2e,float}
\usepackage[parfill]{parskip}
\usepackage{xcolor}
\usepackage[font=normalsize,labelfont=bf]{caption}
\usepackage[bookmarks=true,colorlinks=true,linkcolor=blue,citecolor=red]{hyperref}
\biboptions{numbers,sort&compress}
\makeatletter
\AtBeginDocument{\def\@citecolor{red}}
\makeatother

\journal{Elsevier}

\begin{document}

\begin{frontmatter}



\title{A failure mode dependent continuum damage model for laminated composites with optimized model parameters : Application to curved beams}


\author[1]{Shubham Rai\texorpdfstring{\corref{cor1}}{}}
\ead{shubham.rai@am.iitd.ac.in}
\author[1]{Badri Prasad Patel\texorpdfstring{\corref{cor1}}{}}
\ead{bppatel@am.iitd.ac.in}
\cortext[cor1]{Communicating authors}
\affiliation[1]{organization={Department of Applied Mechanics, Indian Institute of Technology Delhi},
            addressline={Hauz khas}, 
            postcode={110016}, 
            state={New Delhi},
            country={India}}

\begin{abstract}

In this article, a failure mode dependent and thermodynamically consistent continuum damage model with polynomial-based damage hardening functions is proposed for continuum damage modeling of laminated composite panels. The damage model parameters are characterized based on all uniaxial/shear experimental stress-strain curves. Steepest descent optimization algorithm is used to minimize the difference between model predicted and experimental stress-strain curves to get the optimzed model parameters. The fully characterized damage evolution equations are used for damage prediction of a moderately thick laminated composite curved beam modeled using {first-order shear deformation theory}. Finite element method with load control is used to get the non-linear algebraic equations which are solved using Newton Raphson method.{ The developed model is compared with the existing failure mode dependent and failure mode independent damage models}. The results depict the efficacy of the proposed model to capture {non-linearity in the load vs deflection curve due to stiffness degradation and} different damage in tension and compression consistent with uniaxial/shear stress-strain response and strength properties of the material{, respectively.}
\begin{keyword}

{Continuum Damage Mechanics, Failure mode dependent, Finite element method, Fiber reinforced composites, Curved beams}

\end{keyword}

\end{abstract}

\end{frontmatter}

\section{Introduction} 

Brittle fiber-reinforced composites undergo stiffness degradation upon application of load due to the evolution of micro-cracks/voids. This stiffness degradation can be modeled using Continuum Damage Mechanics (CDM). The thermodynamically consistent continuum damage models involve damage hardening functions in terms of an equivalent damage parameter to account for the energy dissipation due to damage. Robbins et al. \cite{robbins2005efficient} investigated the effect of linear and exponential damage hardening functions on damage evolution and uniaxial/shear stress-strain behavior of fiber reinforced composites using failure mode independent damage model proposed by Barbero and De Vivo \cite{barbero2001constitutive}. Gupta and Patel \cite{gupta2012continuum} used the Barbero and De Vivo \cite{barbero2001constitutive} damage model with linear damage hardening function for damage prediction of laminated cylindrical/conical panels. To the best of the authors knowledge, the effect of different polynomial-based damage hardening functions (optimized using all uniaxial/shear stress-strain curves) on the damage evolution of laminated composites has not been investigated in the literature to the best of the author's knowledge. Therefore, in the present work, a thermodynamically consistent failure mode-dependent damage model with polynomial-based damage hardening functions and optimized model parameters is proposed for laminated composites, and its application is demonstrated considering laminated composite curved beams.

\section{Formulation}

\subsection{Failure mode dependent continuum damage model development}
{Fiber-reinforced composites involve different failure mechanisms in different directions (viz. fiber rupture, fiber kinking/micro-buckling, matrix cracking, matrix crushing, delamination, etc.). Failure mode independent damage evolution equations cannot account for these mechanisms accurately. Further, the damage evolution equations can be derived either based on experimental observations (phenomenological damage models) or laws of irreversible thermodynamics. However, the experimental observations to propose damage evolution equations are not always available a priori and involve extensive testing of the material which is cost intensive. }
Therefore, a three-dimensional thermodynamically consistent failure mode-dependent continuum damage model is developed in this section. {The failure modes considered in this paper are fiber rupture, fiber kinking/micro-buckling, matrix cracking, and matrix crushing.} The damage is represented by second-order symmetric damage tensor, which is expressed in principal material direction coordinate space in Voigt notation as $\underline{D}=[D_{11} \ D_{22} \ D_{33} \ D_{12} \ D_{13} \ D_{23}]^T$. The subscripts 1, 2, and 3 represent the principal material coordinates of fiber-reinforced composite material in fiber, in-plane, and out-of-plane transverse to fiber direction, respectively. The fourth-order symmetric compliance tensor for an orthotropic composite material depending on the damage variables (based on \cite{matzenmiller1995constitutive}) is given in contracted notation as : 

\begin{align}\label{compliance_mat}
    \underline{\underline{{H}}}(\underline{D})=
    \begin{bmatrix}
        \frac{1}{E_1(1-D_{11})} & -\frac{\nu_{21}}{E_2} & -\frac{\nu_{31}}{E_3} & 0 & 0 & 0\\
        -\frac{\nu_{21}}{E_2} & \frac{1}{E_2(1-D_{22})} & -\frac{\nu_{32}}{E_3} & 0 & 0 & 0\\
        -\frac{\nu_{31}}{E_3} & -\frac{\nu_{32}}{E_3} & \frac{1}{E_3(1-D_{33})} & 0 & 0 & 0\\
        0 & 0 & 0 & \frac{1}{G_{12}(1-D_{12})} & 0 & 0\\
        0 & 0 & 0 & 0 & \frac{1}{G_{13}(1-D_{13})} & 0\\
        0 & 0 & 0 & 0 & 0 & \frac{1}{G_{23}(1-D_{23})}\\
    \end{bmatrix}
\end{align}

where $E_1, E_2$ and $E_3$ are Young's moduli of elasticity in fiber, in-plane, and out-of-plane transverse to fiber directions, respectively. $G_{12}$ is in-plane and $G_{13}$, $G_{23}$ are out-of plane shear moduli of elasticity.The elastic constitutive relation for a lamina with evolving damage is given by $\underline{\sigma}=\underline{\underline{C}}(\underline{D})\underline{\varepsilon}$ , where $\underline{\underline{C}}=\underline{\underline{{H}}}^{-1}(\underline{D})$ is the elastic constitutive tensor depending on damage variable $\underline{D}$ and $\underline{\sigma}=[\sigma_{11} \ \sigma_{22} \ \sigma_{33} \ \sigma_{12} \ \sigma_{13} \ \sigma_{23}]^T$, $\underline{\varepsilon}=[\varepsilon_{11} \ \varepsilon_{22} \ \varepsilon_{33} \ 2\varepsilon_{12} \ 2\varepsilon_{13} \ 2\varepsilon_{23}]^T$ are stress and strain tensors expressed in Voigt notation, respectively.

The Helmholtz free energy density of a composite lamina with evolving damage is split into an elastic recoverable strain energy density $\phi_{e}$ and a dissipative energy density $\pi$ depending on an internal state variable $\beta$, also known as damage hardening variable. It is expressed as  \cite{robbins2005efficient,barbero2001constitutive} : 

\begin{equation}\label{psi}
    \psi(\underline\varepsilon_e,\underline{D},\beta)=\frac{1}{\rho}(\phi_{e}(\underline\varepsilon_e,\underline{D}) + \pi(\beta))
\end{equation}

\begin{figure}[h!]
    \centering
    \includegraphics[scale=0.5]{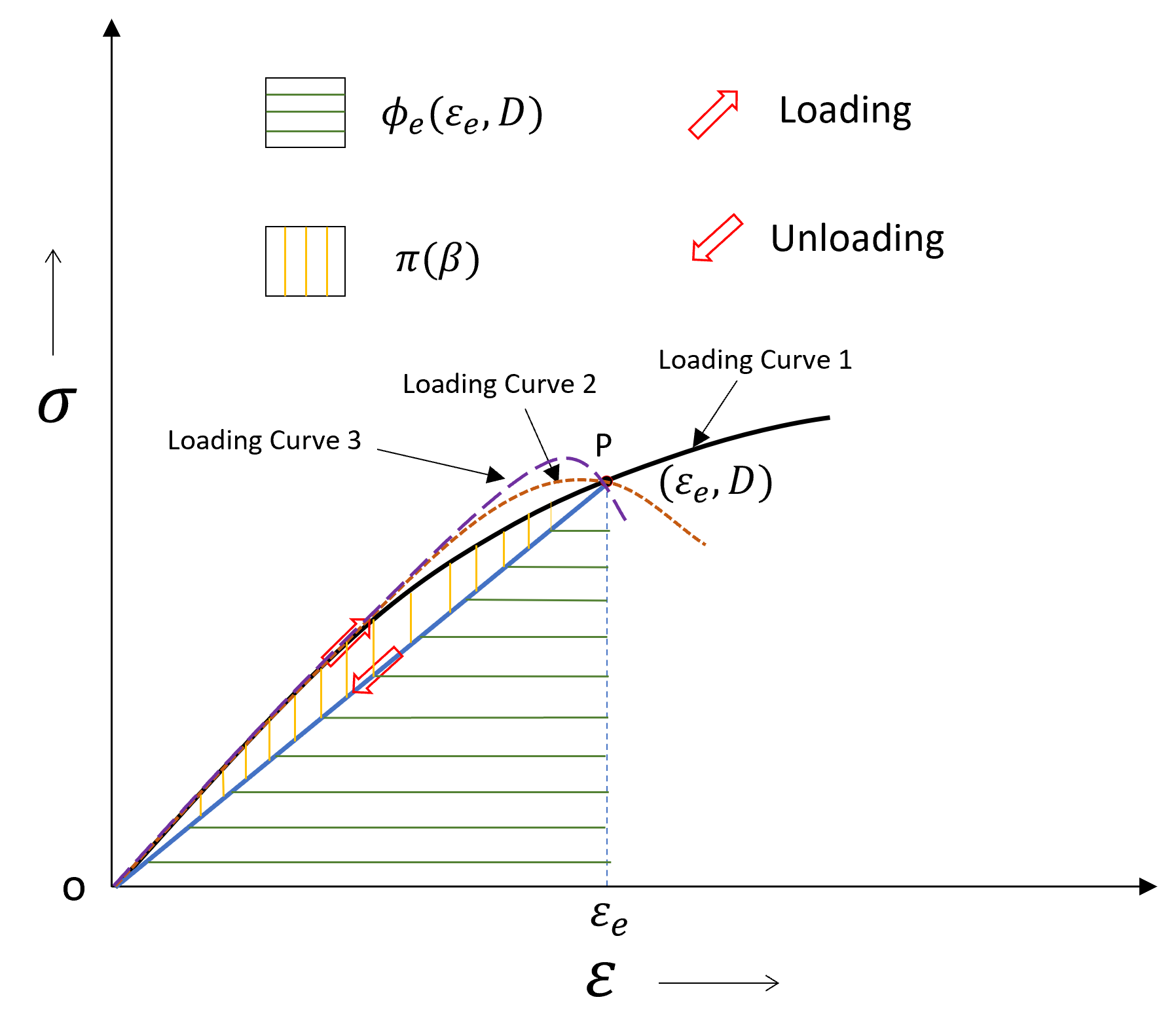}
    \caption{{Representation of recoverable ($\phi_e$) and dissipative ($\pi$) energy densities}}
    \label{fig:phi_pi}
\end{figure}

{ For the same state of damage ($\underline{D}$) and elastic strain ($\underline{\varepsilon}_e$), there can be different amounts of energy dissipation depending on the micro-structure or grain structure of the material} (Fig.~\ref{fig:phi_pi}). {Therefore, an independent variable $\beta$ is required to account for the amount of energy dissipation. }

\subsubsection{Damage evolution equations}

Damage evolution equations are derived based on laws of irreversible thermodynamics {without resorting to experimental observations a priori}. The second law of thermodynamics invoking isothermal conditions, elastic deformation with evolving damage and zero heat flux exchange, is given in terms of Clausius-Duhem inequality as \cite{hao1985anisotropic} : 

\begin{equation}\label{clausius-duhem}
    \left(\underline{\sigma}-\frac{\partial{(\rho\psi)}}{\partial{\underline\varepsilon_e}}\right) \dot{\underline\varepsilon}_e - \frac{\partial(\rho\psi)}{\partial{\underline{D}}}\dot{\underline{D}} - \frac{\partial(\rho\psi)}{\partial{\beta}}\dot{\beta} \geq0
\end{equation}

The thermodynamic damage driving force ($\underline{Y}$) conjugate  to damage variable $\underline{D}$ and damage hardening function ($\gamma$) conjugate to damage hardening variable $\beta$ are defined as :

\begin{equation}\label{eq: Y}
    \underline{Y}=-\frac{\partial{(\rho\psi)}}{\partial{\underline{D}}}=-\frac{\partial{\phi_e}}{\partial{\underline{D}}}=\frac{\partial{\chi}}{\partial{\underline{D}}}=\frac{1}{2}\underline{\sigma} : \frac{\partial\underline{\underline{H}}(\underline{D})}{\partial\underline{D}}:\underline{\sigma}
\end{equation}

\begin{equation}\label{gamma}
    \gamma(\beta)=\frac{\partial{(\rho\psi)}}{\partial{\beta}}=\frac{\partial{(\pi)}}{\partial{\beta}}
\end{equation}

where $\chi(\underline{\sigma},\underline{D})$ is complementary energy density.

Substituting Eqs.~\ref{eq: Y},~\ref{gamma} in Eq.~\ref{clausius-duhem} and considering the arbitrary nature of $\dot{\varepsilon}_e$ gives : 

\begin{equation}\label{sigma_thermo}
    \underline{\sigma}=\frac{\partial{(\rho\psi)}}{\partial{\underline\varepsilon_e}}
\end{equation}

\begin{equation}\label{diss_thermo}
    \Phi(\underline{Y},\gamma)=Y_i\dot{D}_i-\gamma\dot{\beta} \geq0
\end{equation}

where $\Phi(\underline{Y},\gamma)$ is known as damage dissipation power density. The dissipation power density is considered to be failure mode dependent and the principle of maximum dissipation (subject to consistency condition $f_m(\underline{Y})=0$, where $m$ represents damage mode) is used to arrive at failure mode dependent damage evolution equations in $m^{th}$ mode.$f_m$ denotes the loading criteria function in $m^{th}$ mode in thermodynamic damage driving force ($\underline{Y}$) space. Lagrange multiplier approach is used to maximize $\Phi_m(\underline{Y},\gamma_m)$ subject to $f_m(\underline{Y})=0$, where the subscript $m$=$1:ft$,$2:fc$,$3:mt$ and $4:mc$ for fiber tension, fiber compression, matrix tension and matrix compression mode, respectively. The unconstrained objective function is expressed as :

\begin{equation}\label{mode_diss_thermo}
    \Phi^*_m(\underline{Y},\gamma_m)=\underline{Y}.\dot{\underline{D}_m}-\gamma_m\dot{\beta_m} -\lambda f_m
\end{equation}
where $\lambda$ is Lagrange multiplier and repeated index $m$ does not imply summation. The first-order necessary condition for maxima of $\phi_m^*$ give :

\begin{align}\label{dam_ev_mode}
    \dot{\underline{D}}_m=\dot{\beta}_m\frac{\partial{f_m}}{\partial{\underline{Y}}}
\end{align}

The total increment in damage is the summation of incremental damage variables from each mode as:

\begin{equation}\label{dam_ev_total}
    \dot{\underline{D}}=\sum_{m=1}^4 \dot{\beta}_m\frac{\partial{f_m}}{\partial{\underline{Y}}}
\end{equation}


\subsection{Loading criteria functions}

The loading criteria functions (based on \cite{hashin1980failure,matzenmiller1995constitutive,sosa2013modelling}), with different damage hardening functions and damage hardening variable in each mode, are expressed in four modes of damage in stress space as : 

Fiber Tension ($\sigma_{11}>0$) :
\begin{align}\label{f1}
     f_1=\left(\frac{\sigma_{11}}{(1-D_{11})X_t}\right)^2 + \left(\frac{\sigma_{12}}{(1-D_{12})S_a}\right)^2 + \left(\frac{\sigma_{13}}{(1-D_{13})S_a}\right)^2 - \gamma_{1}(\beta_{1}) = \underline{\sigma}^T\underline{\underline{F_1}} \underline{\sigma}- \gamma_{1}(\beta_{1})   
\end{align}

Fiber Compression ($\sigma_{11}<0$) :
\begin{align}\label{f2}
    f_2=\left(\frac{\sigma_{11}}{(1-D_{11})X_c}\right)^2 - \gamma_{2}(\beta_{2})=\underline{\sigma}^T\underline{\underline{F_2}} \underline{\sigma} - \gamma_{2}(\beta_{2}) 
\end{align}

Matrix Tension $(\sigma_{22}+\sigma_{33})>0$ :
\begin{align}\label{f3}
\begin{split}
      f_3=\left(\frac{\sigma_{22}}{(1-D_{22})Y_t}\right)^2 + \left(\frac{\sigma_{33}}{(1-D_{33})Z_t}\right)^2 + \left(\frac{\sigma_{12}}{(1-D_{12})S_a}\right)^2 + \left(\frac{\sigma_{13}}{(1-D_{13})S_a}\right)^2 + \\
    \left(\frac{\sigma_{23}}{(1-D_{23})S_t}\right)^2 - \gamma_{3}(\beta_{3})=\underline{\sigma}^T\underline{\underline{F_3}} \underline{\sigma} - \gamma_{3}(\beta_{3})  
\end{split}
\end{align}

Matrix Compression $(\sigma_{22}+\sigma_{33})<0$ :
\begin{align}\label{f4}
\begin{split}
      f_4=\left(\frac{\sigma_{22}}{(1-D_{22})Y_c}\right)^2 + \left(\frac{\sigma_{33}}{(1-D_{33})Z_c}\right)^2 + \left(\frac{\sigma_{12}}{(1-D_{12})S_a}\right)^2 + \left(\frac{\sigma_{13}}{(1-D_{13})S_a}\right)^2 + \\
    \left(\frac{\sigma_{23}}{(1-D_{23})S_t}\right)^2 - \gamma_{4}(\beta_{4})=\underline{\sigma}^T\underline{\underline{F_4}} \underline{\sigma}- \gamma_{4}(\beta_{4})   
\end{split}
\end{align}

where $X_t, X_c$ are strength in fiber tension and compression, respectively, $Y_t,Y_c$ are strength in in-plane transverse to fiber tension and compression, respectively, $Z_t,Z_c$ are strength in out-of-plane transverse to fiber tension and compression, respectively. $S_a=S_{12}=S_{13}$ is the axial shear strength of the lamina and $S_t=S_{23}$ is the transverse shear strength of the lamina. Tensors $\underline{\underline{F_1}}, \underline{\underline{F_2}}, \underline{\underline{F_3}}$ and $\underline{\underline{F_4}}$ are defined in appendix. The thermodynamic damage driving force in eq.~\ref{eq: Y} in stress tensor $(\underline{\sigma})$ and damage variable tensor $(\underline{D})$ space is written as :
\begin{equation}\label{Y_stress}
\begin{split}
    \underline{Y}(\underline{\sigma},\underline{D})&=[Y_{11},Y_{22},Y_{33},Y_{12},Y_{13},Y_{23}]^T \\
    &=[\frac{\sigma_{11}^2}{2(1-D_{11})^2E_{1}}, \frac{\sigma_{22}^2}{2(1-D_{22})^2E_{2}}, \frac{\sigma_{33}^2}{2(1-D_{33})^2E_{3}}, \frac{\sigma_{12}^2}{2(1-D_{12})^2G_{12}},\frac{\sigma_{13}^2}{2(1-D_{13})^2G_{13}},\\ & \hspace{0.5 cm} \frac{\sigma_{23}^2}{2(1-D_{23})^2G_{23}} ]
\end{split}
\end{equation}

The loading criteria functions $f_m({\underline{\sigma}})$ are converted from stress space to thermodynamic damage driving force space $f_m({\underline{Y}})$ using Eq.~\ref{Y_stress} as :

\begin{align}\label{f1_Y}
f_1=\frac{2Y_{11}E_1}{X_t^2} + \frac{2Y_{12}G_{12}}{S_a^2} + \frac{2Y_{13}G_{13}}{S_a^2} - \gamma_1(\beta_1)
\end{align}

\begin{align}\label{f2_Y}
f_2=\frac{2Y_{11}E_1}{X_c^2} - \gamma_2(\beta_2)
\end{align}

\begin{align}\label{f3_Y}
f_3=\frac{2Y_{22}E_2}{Y_t^2} + \frac{2Y_{33}E_3}{Z_t^2} + \frac{2Y_{12}G_{12}}{S_a^2} + \frac{2Y_{13}G_{13}}{S_a^2} + \frac{2Y_{23}G_{23}}{S_t^2} - \gamma_3(\beta_3)
\end{align}

\begin{align}\label{f4_Y}
f_4=\frac{2Y_{22}E_2}{Y_c^2} + \frac{2Y_{33}E_3}{Z_c^2} + \frac{2Y_{12}G_{12}}{S_a^2} + \frac{2Y_{13}G_{13}}{S_a^2} + \frac{2Y_{23}G_{23}}{S_t^2} - \gamma_4(\beta_4)
\end{align}

Finally, the thermodynamic damage driving force in stress space $\underline{Y}(\underline{\sigma},\underline{D})$ is expressed in strain space, i.e. $\underline{Y}(\underline{\varepsilon},\underline{D})$, using the constitutive relation $\underline{\sigma}=(\underline{\underline{C}}(\underline{D}))\underline{\varepsilon}$.

\subsubsection{Damage hardening functions}

Due to different mechanisms of damage evolution in different directions in fiber-reinforced composite materials, there is a need to propose different damage hardening variables ($\beta_m$) in each mode $m$. Polynomial-based damage hardening functions are selected based on the minimum difference between the model predicted and experimental uniaxial/shear stress-strain curves in the numerical characterization of damage model parameters. Three types of damage hardening functions are taken, i.e., linear, quadratic, and cubic polynomials, with damage hardening function parameters dependent on damage mode $m$. The damage hardening functions are expressed as :

\begin{equation}\label{gamma_m_cubic}
    \gamma_m(\beta_m)=C_{1m}\beta_{m}+C_{2m}{\beta_{m}}^2+C_{3m}{\beta_{m}}^3
\end{equation}

For linear damage hardening :  $C_{1m}>0$, $C_{2m}=0$ and $C_{3m}=0$

For quadratic damage hardening : $C_{1m}>0$, $C_{2m}>0$ and $C_{3m}=0$

For cubic damage hardening : $C_{1m}>0$, $C_{2m}>0$ and $C_{3m}>0$

The damage evolution equations are solved iteratively using the value of strain $(\underline{\varepsilon})$ at the current iteration and the damage variable value obtained from the previous iteration to get the damage variable at the current iteration by satisfying loading criteria ($f_m(\underline{Y})=0$) in each mode.





 \subsection{Damage model parameter characterization of the developed failure mode dependent damage model}
 
The linear, quadratic, and cubic polynomial-based damage hardening functions are considered for optimization of model parameters of the proposed three-dimensional thermodynamically consistent failure mode dependent continuum damage model (with damage evolution equations involving six damage variables $D_{11}, D_{22}, D_{33}$, $ D_{12}, D_{13}$ and $D_{23}$, derived based on second law of thermodynamics by simultaneously satisfying the loading criteria by using functions derived based on Hashin (1980) \cite{hashin1980failure} failure criteria) to minimize the difference between model predicted and experimental uniaxial/shear stress-strain curves. Steepest descent algorithm (for optimization) coupled with Newton-Raphson iterative procedure (for solving damage evolution equations by satisfying loading criteria) is used to arrive at an optimized set of damage model parameters $\underline{\mathcal{P}}=[C_{1m} \ C_{2m} \ C_{3m}]$. The loss/objective function based on the squared difference between the model predicted and experimental uniaxial/shear stress-strain curve is given as:
\vspace{-5pt}
\begin{equation}\label{mode_loss}
    \pounds_{m_a}(\underline{\mathcal{P}}) =\sum_{j=1}^{n_{m_a}} \sum_{i=1}^{n} \left({\frac{ \ {\sigma}^i_{(j)exp} - {\sigma}^i_{(j)model}(\underline{\mathcal{P}})\ }{S_j}}\right)^2
\end{equation}
where $i$ and $j$ represent summation over experimental data points and number of active stress-strain curves, respectively. $S_j$ represents strength value in $j^{th}$ uniaxial/shear stress-strain curve. The summation of the active mode dependent loss function over all the possible combinations of active modes ($m_a$) gives the total loss function $\pounds(\underline{\mathcal{P}})$.

\begin{figure}[h!]
\begin{subfigure}{0.5\textwidth}
\includegraphics[width=\textwidth]{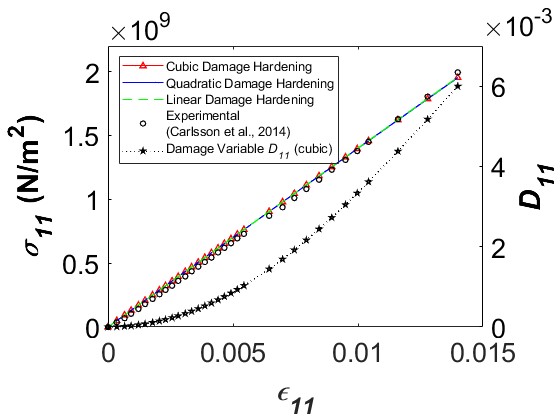}
\label{fig:s33_e33_ftmt}
\end{subfigure}
\begin{subfigure}{0.5\textwidth}
\includegraphics[width=\textwidth]{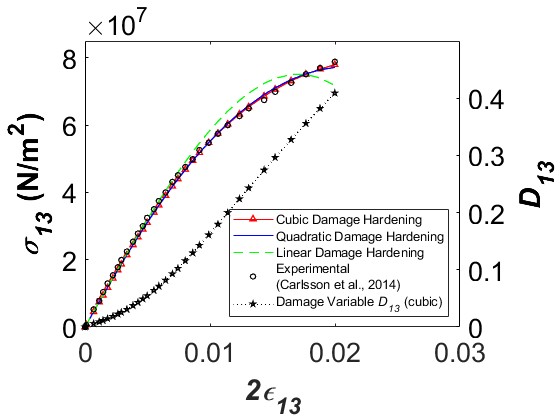}
\label{fig:s13_e13_ftmt}
\end{subfigure}
\caption{{Experimental versus optimized model predicted stress-strain curves and damage variable vs strain curves corresponding to linear, quadratic and cubic damage hardening for the $ft-mt$ active set of damage modes.} }
\label{fig:uniaxial_ftmt}
\end{figure}

\begin{figure}[h!]
\begin{subfigure}{0.5\textwidth}
\includegraphics[width=\textwidth]{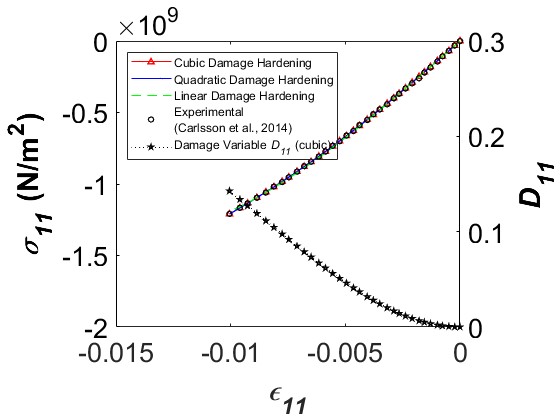}
\label{fig:s33_e33_fcmc}
\end{subfigure}
\begin{subfigure}{0.5\textwidth}
\includegraphics[width=\textwidth]{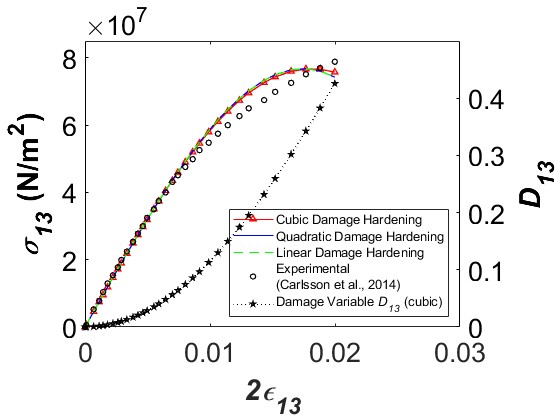}
\label{fig:s13_e13_fcmc}
\end{subfigure}
\caption{{Experimental versus optimized model predicted stress-strain curves and damage variable vs strain curves corresponding to linear, quadratic and cubic damage hardening for the $fc-mc$ active set of damage modes.} }
\label{fig:uniaxial_fcmc}
\end{figure}

AS4/3501-6 fiber reinforced brittle composite with material properties taken from \cite{kaddour2013mechanical} and \cite{carlsson2014experimental} (Table \ref{Material Properties}) is considered for the analysis. The model predicted, and experimental in-plane normal and out-of-plane shear stress vs strain curves with linear, quadratic, and cubic damage hardening for the damage modes active in fiber tension - matrix tension ($ft-mt$) and fiber compression - matrix compression ($fc-mc$) are shown in Fig.~\ref{fig:uniaxial_ftmt} and Fig.~\ref{fig:uniaxial_fcmc}, respectively. These curves depict that the damage hardening function with a polynomial up to cubic terms gives the most accurate results. Damage variable vs. strain curves corresponding to cubic damage hardening (Fig.~\ref{fig:uniaxial_ftmt}) depicts more damage in the out-of-plane shear direction.\\

The optimized damage model parameters and active mode dependent loss function values for cubic damage hardening are given as:\\

$C_{1ft}=0.1027 \times 10^{-4}$, $C_{2ft}= 0.3354 \times 10^{-14}$, $C_{3ft}=0.2105\times 10^{-15}$ \\
\ \ $C_{1mt}=0.5966 \times 10^{-9}$, $C_{2mt}=0.1203\times 10^{-14}$, $C_{3mt}= 0.2263 \times 10^{-12}$ \\
$\pounds_{ft-mt}(\underline{\mathcal{P}})=0.03442$ \\

\ \ $C_{1fc}=0.1276\times 10^{-5}$, $C_{2fc}=0.5388 \times 10^{-12}$, $C_{3fc}= 0.4125 \times 10^{-18}$ \\
\ \ $C_{1mc}=0.1274\times 10^{-4}$, $C_{2mc}=0.2904 \times 10^{-14}$, $C_{3mc}= 0.2735 \times 10^{-16}$ \\
$\pounds_{fc-mc}(\underline{\mathcal{P}})=0.236149$

\begin{table}[h!]
    \centering
    \   \begin{tabular}{  m{3cm}  m{2cm}  c  m{2cm} }
    \hline
           Elastic Properties &   Values            & Strength Properties   & Values \\
    \hline
        \hspace{0.6cm}   $E_1$                & 140.4 GPa         & $X_t$                 & 1980 MPa  \\
        \hspace{0.6cm}   $E_2=E_3$            & 11 \ \ \ \ GPa    & $X_c$                 & 1200 MPa  \\
        \hspace{0.6cm}   $G_{12}=G_{13}$      & 6.60   \ \  GPa   & $Y_t$                 & 53  \ \ \ MPa  \\
        \hspace{0.6cm}     $G_{23}$             & 3.62   \ \ GPa    & $Y_c$                 & 200 \ \ MPa  \\
        \hspace{0.6cm}    $\nu_{12}=\nu_{13}$  & 0.28              & $S_a=S_{12}=S_{13}$   & 79  \ \ \ MPa   \\
        \hspace{0.6cm}     $\nu_{23}$           & 0.52              & $S_t=S_{23}$          & 55 \ \ \  MPa   \\
    \hline
    \end{tabular}
    \caption{Material Properties for AS4/3501-6 fiber-reinforced composite \cite{kaddour2013mechanical,carlsson2014experimental}}
    \label{Material Properties}
\end{table}

\section{Finite Element Formulation}
\begin{figure}[h!]
    \centering
    \includegraphics[scale=0.7]{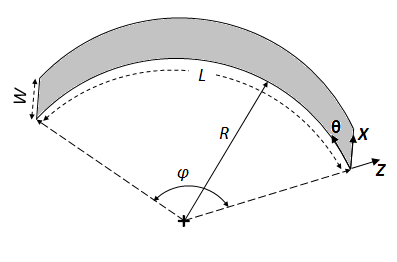}
    \caption{{Geometry and coordinate system of a curved panel} }
    \label{fig:curved_panel}
\end{figure}

{First-order shear deformation theory is used for modeling a moderately thick laminated composite curved panel subjected to transverse loading}. The geometry and coordinate system of the panel are shown in Fig.~\ref{fig:curved_panel} where $R$ is the radius of curvature of the panel, $L$ is the length of the panel along $\theta$ coordinate and $\varphi$ is the sector angle of the panel such that $L=R$ $\varphi$.

{The displacements $u$, $v$ and $w$ along the $x$, $\theta$ and $z$ directions, respectively are given as} : 

\begin{equation}
    \begin{gathered}
        u(x,\theta,z)= u_o(x,\theta) +z\psi_x(x,\theta) \\
        v(x,\theta,z)= v_o(x,\theta) +z\psi_{\theta}(x,\theta) \\
        w(x,\theta,z)=w_o(x,\theta)
    \end{gathered}
\end{equation}

{where $u_o$, $v_o$ and $w_o$ are mid-plane displacements. $\psi_x$ and $\psi_{\theta}$ are the rotations of shell normal about $\theta$ and negative $x$ axes, respectively.}

The strain displacement relationship based on the assumptions of (i) small strains and rotations, and (ii) thin shells ($1+z/R \approx 1$) are given as :

\begin{equation}\label{Eq: strain_disp rel}
   \begin{gathered}
        \{\varepsilon\}=
        \begin{Bmatrix}
            \varepsilon_{x x} \\[0.1cm]
            \varepsilon_{\theta \theta} \\[0.1cm]
            \gamma_{x \theta} \\[0.1cm]
            \gamma_{x z} \\[0.1cm]
            \gamma_{\theta z}
        \end{Bmatrix}=
        \begin{Bmatrix}
            \frac{\partial{u_o}}{\partial{x}} \\[0.1cm]
            \frac{1}{R}\frac{\partial{v_o}}{ \partial{\theta}}  + \frac{w_o}{R} \\[0.1cm]
            \frac{1}{R}\frac{\partial{u_o}}{\partial{\theta}} + \frac{\partial{v_o}}{\partial{x}} \\[0.1cm]
            0 \\[0.1cm]
            0 
        \end{Bmatrix} +  
        \begin{Bmatrix}
            z\frac{\partial{\psi_x}}{\partial{x}} \\[0.1cm]
            \frac{z}{R}\frac{\partial{\psi_{\theta}}}{\partial{\theta}}\\[0.1cm]
            \frac{z}{R}\frac{\partial{\psi_x}}{\partial{\theta}} +z\frac{\partial{\psi_{\theta}}}{\partial{x}}\\[0.1cm]
            \psi_x + \frac{\partial{w_o}}{\partial{x}}\\[0.1cm]
            \psi_{\theta} + \frac{1}{R}\frac{\partial{w_o}}{\partial{\theta}} - \frac{v_o}{R}
        \end{Bmatrix}
    \end{gathered} 
\end{equation}

{The whole geometry is discretized into $C^0$ continuous finite elements. All the numerical integrations are performed using Gauss Quadrature numerical integration scheme with 3$\times$3 Gauss points for integration with respect to $x$ and $\theta$, and $5$ Gauss points in each layer for integration with respect to $z$.} The total load is applied in a number of steps to ensure convergence of nodal degrees of freedom within each load step. Every load step involves global iterations in which the damage evolution equations are solved iteratively at each Gauss point using the strain values at the current global iteration by satisfying the loading criteria $f_m=0$.

\section{Results and discussion}

The above formulation is used for damage prediction in a laminated composite curved beam {($L/W=$1000)} with a lamination scheme as 90/90/90/90 with respect to $x$ coordinate, i.e., the direction of fibers in all four layers are along $\theta$ direction. The dimensions of the beam are : $\varphi=0.4$ rad, $R=4m$ i.e. $L=1.6m$, $L/h=10$ where $h$ is the thickness of the beam. The material properties are taken from Table.~\ref{Material Properties}. Both sides of the beam are clamped to restrict all the displacement and rotation degrees of freedom, and the beam is subjected to uniformly distributed transverse load in the positive $z$ direction (radially outward).

The present failure mode dependent damage model is compared with the failure mode independent damage model of Barbero and De Vivo \cite{barbero2001constitutive} with constitutive relations taken from \cite{robbins2005efficient} and {an existing uncoupled form of phenomenological failure mode dependent damage model} \cite{pan2022use} {with compliance matrix given by} Eq.~\ref{compliance_mat} . {A mesh size of 10$\times$1 is taken for the present and Barbero and De Vivo damage models. A mesh size of 1000$\times$1 is taken for the existing failure mode-dependent damage model based prediction.} The damage model parameters for Barbero and De Vivo damage model are taken from Rai and Patel \cite{rai2023failure} and are given in Table~\ref{table : barb_optm}. Babero's associative damage model ($H_1=H_2=H_3=0$) with linear damage hardening is considered for damage prediction of the curved beam.

{The damage evolution equations of the existing failure mode dependent damage model (based on Pan et al.} \cite{pan2022use}) {are given as :}

\begin{equation}\label{Eq : exponential damage evolution equation}
    d_i^j=1-\frac{1}{f_i^j}\exp{\frac{(1-f_i^j) \varepsilon_{fi}^j S_i^j l_c}{G_{ci}^j}}
\end{equation} 

{where $i=f$ (fiber), $m$ (matrix) and $j=t$ (tension), $c$ (compression), $S$ represents the strength,$f_i^j$ is the failure criteria function, $\varepsilon_f$ represents the failure strain, $l_c$ is the characteristic length of the element taken as square root of integration point area} \cite{maimi2007bcontinuum,lapczyk2007progressive} {( $l_c$ = 0.0005333 for the considered mesh size) and $G_c$ is the fracture energy. Also, $d_f^j$ = $D_{11}$, $d_m^j$ = $D_{22},D_{33}$, $D_{12}=1-(1-D_{11})(1-D_{22})$, $D_{13}=1-(1-D_{11})(1-D_{33})$ and $D_{23}=1-(1-D_{22})(1-D_{33})$. The strength properties required for this model are taken from} Table~\ref{Material Properties}, {fracture toughness values are taken from} \cite{kim2013composite} {and are given as: $G_{cf}^t=$91,600 J/m$^2$, $G_{cf}^c=$ 79,900 J/m$^2$, $G_{cm}^t=$ 220 J/m$^2$ and $G_{cm}^c=$ 760 J/m$^2$, failure strains are taken from} \cite{kaddour2013mechanical} {and are given as : $\varepsilon_{ff}^t=0.014$, $\varepsilon_{ff}^c=0.01$, $\varepsilon_{fm}^t=0.0055$ and $\varepsilon_{fm}^c=0.02$.}  

\begin{table}[h!]
    \centering
    \begin{tabular}{ c  c  c  c} 
    \hline
     $\pounds, \underline{\mathcal{P}}$ \ \ & Linear Hardening  & $\underline{\mathcal{P}}$ & Values\\
     \hline
     $\pounds$ &  0.021388 &    $J_{11}$ & $1.7588\times10^{-15}$ \\
     $C_1$     & \ \ $5.8573\times10^{-7}$          & $J_{22}=J_{33}$ & $1.8263\times10^{-13}$\\
     $C_2$     & 0        & $H_{1}$ & $1.0302\times10^{-8}$ \\
     $C_3$     & 0        & $H_{2}=H_{3}$ & $-7.8913\times10^{-9}$\\
     \hline
    \end{tabular}
    \caption{Damage model parameters and loss function values for failure mode independent damage model \cite{barbero2001constitutive} for AS4/3501-6 composite material \cite{rai2023failure}.}
    \label{table : barb_optm}
\end{table}
\begin{figure}[h!]
    \centering
    \includegraphics[scale=0.8]{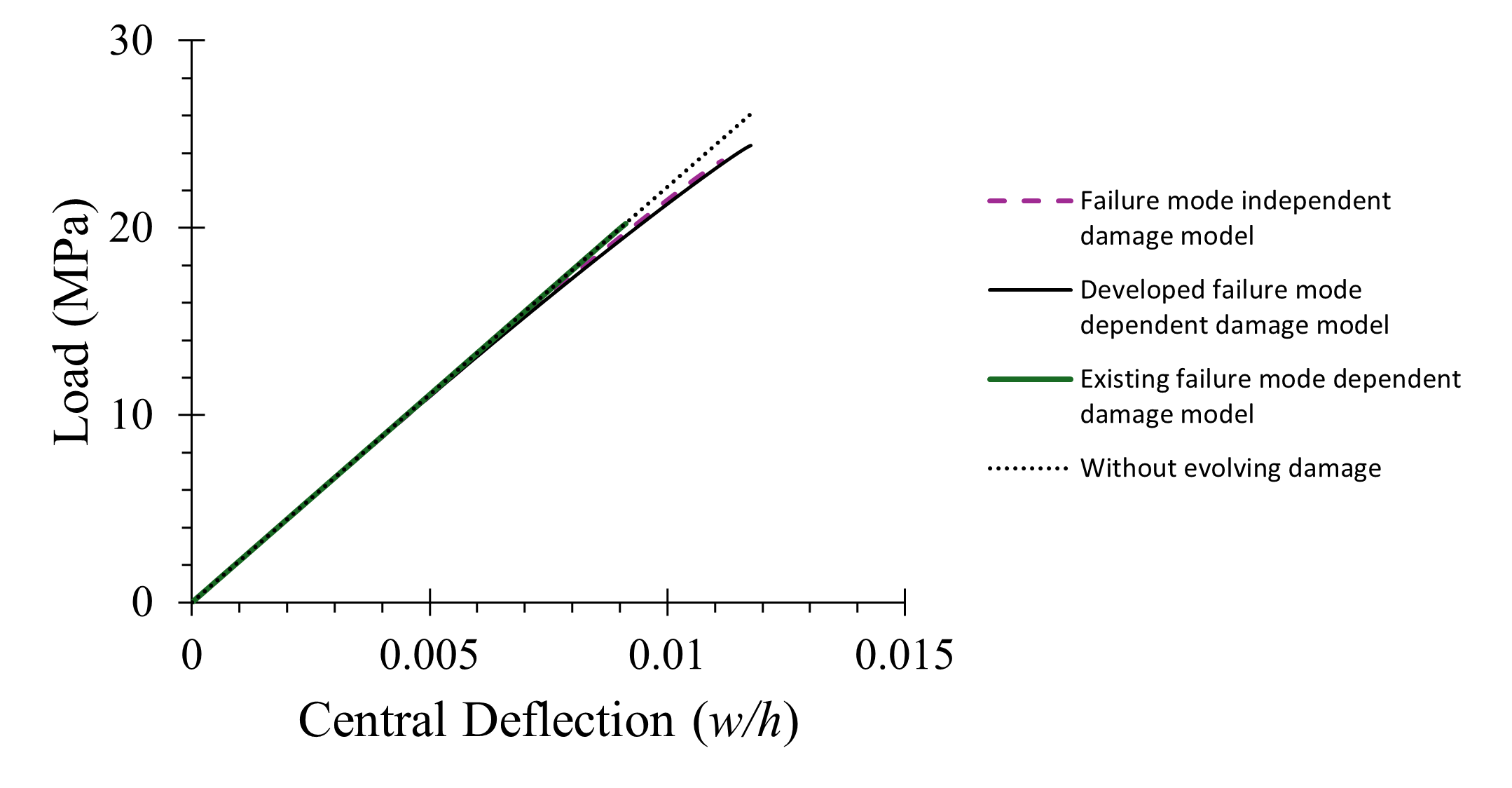}
    \caption{{Load vs central deflection of the curved laminated composite beam (90/90/90/90) for the Present damage model, and Barbero and De Vivo} \cite{barbero2001constitutive} {associative damage model and an exisiting failure mode dependent damage model} \cite{pan2022use}.}
    \label{fig:ldc}
\end{figure}

The load vs central deflection for the laminated composite curved beam is shown in Fig.~\ref{fig:ldc} for the proposed damage model with cubic damage hardening, the Barbero and De Vivo \cite{barbero2001constitutive} associative damage model with linear damage hardening { and the existing failure mode dependent phenomenological damage model} \cite{pan2022use}. The figure depicts the stiffness degradation of laminated composite curved beam due to damage evolution with failure loads (failure load is the load when either the damage variable reaches one or there is convergence failure of the employed load control algorithm) {as 24.39 MPa, 23.56 MPa and 20.24 MPa for the proposed damage model, Barbero and De Vivo} \cite{barbero2001constitutive} {associative damage model and existing failure mode dependent damage model, respectively.} 

{The existing failure mode dependent damage model} \cite{pan2022use} {involves damage evolution when failure criteria function $f_i^j$ at any Gauss point is greater than 1 (which is satisfied after 19.3 MPa for the present geometry, loading, and boundary conditions). Also, the damage variable vs strain curve is concave downward for the existing failure mode-dependent damage model, leading to a rapid damage evolution at the beginning. Therefore, the load vs central deflection curve is almost linear for this model.}

{Significant non-linearity in load vs central deflection curves is predicted by the developed failure mode dependent damage model and Barbero and De Vivo} \cite{barbero2001constitutive} {failure mode independent damage model.} Although the failure loads for both the damage models are fairly close, there is a significant difference between the distribution of damage variables $D_1,D_3$ (Barbero and De Vivo model \cite{barbero2001constitutive}) and $D_{11},D_{33}$ (present developed model).

\begin{figure}[h!]
\begin{subfigure}{0.5\textwidth}
\includegraphics[width=\textwidth]{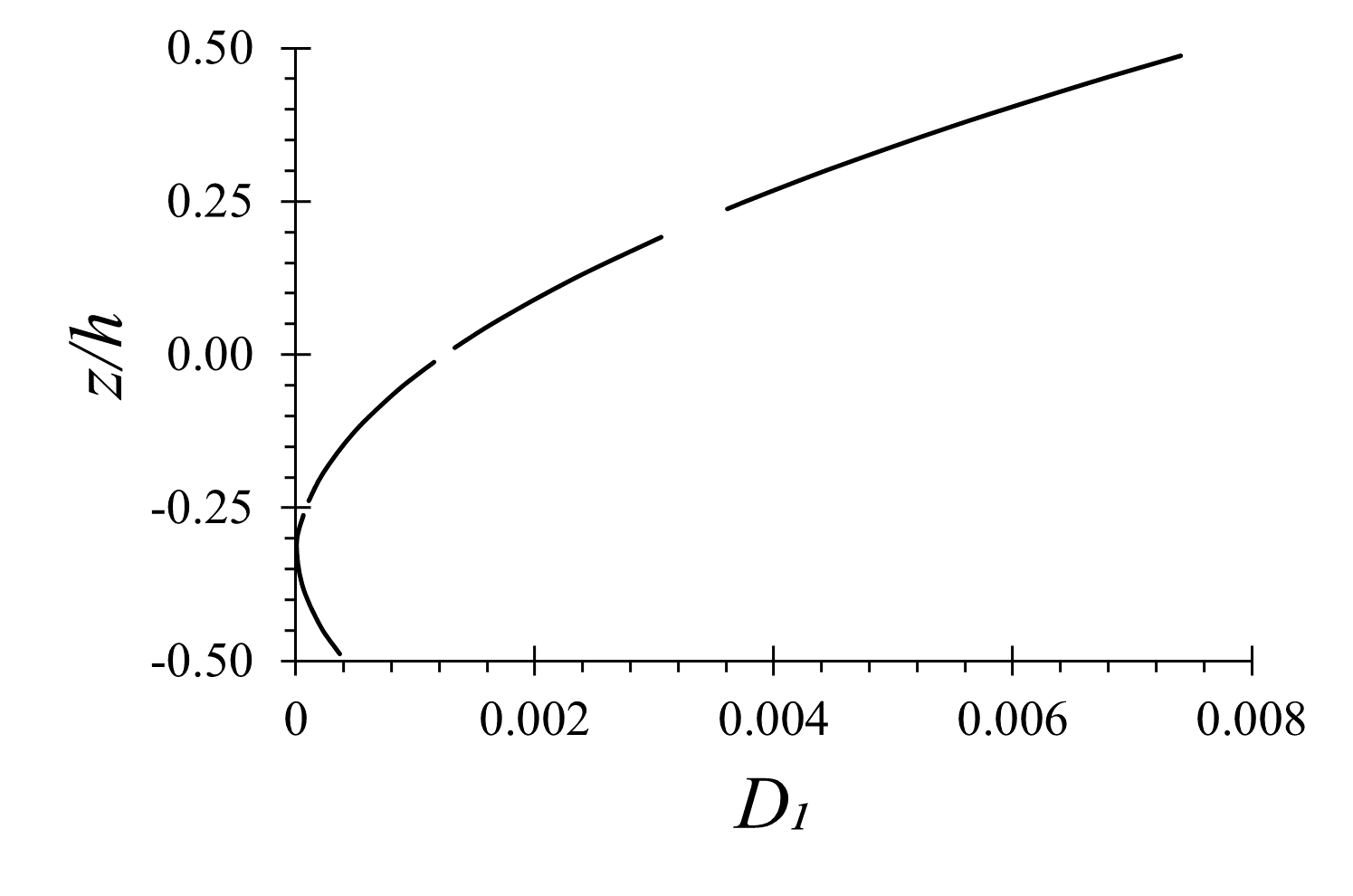}
\caption{}
\label{fig:d1c_barb}
\end{subfigure}
\begin{subfigure}{0.5\textwidth}
\includegraphics[width=\textwidth]{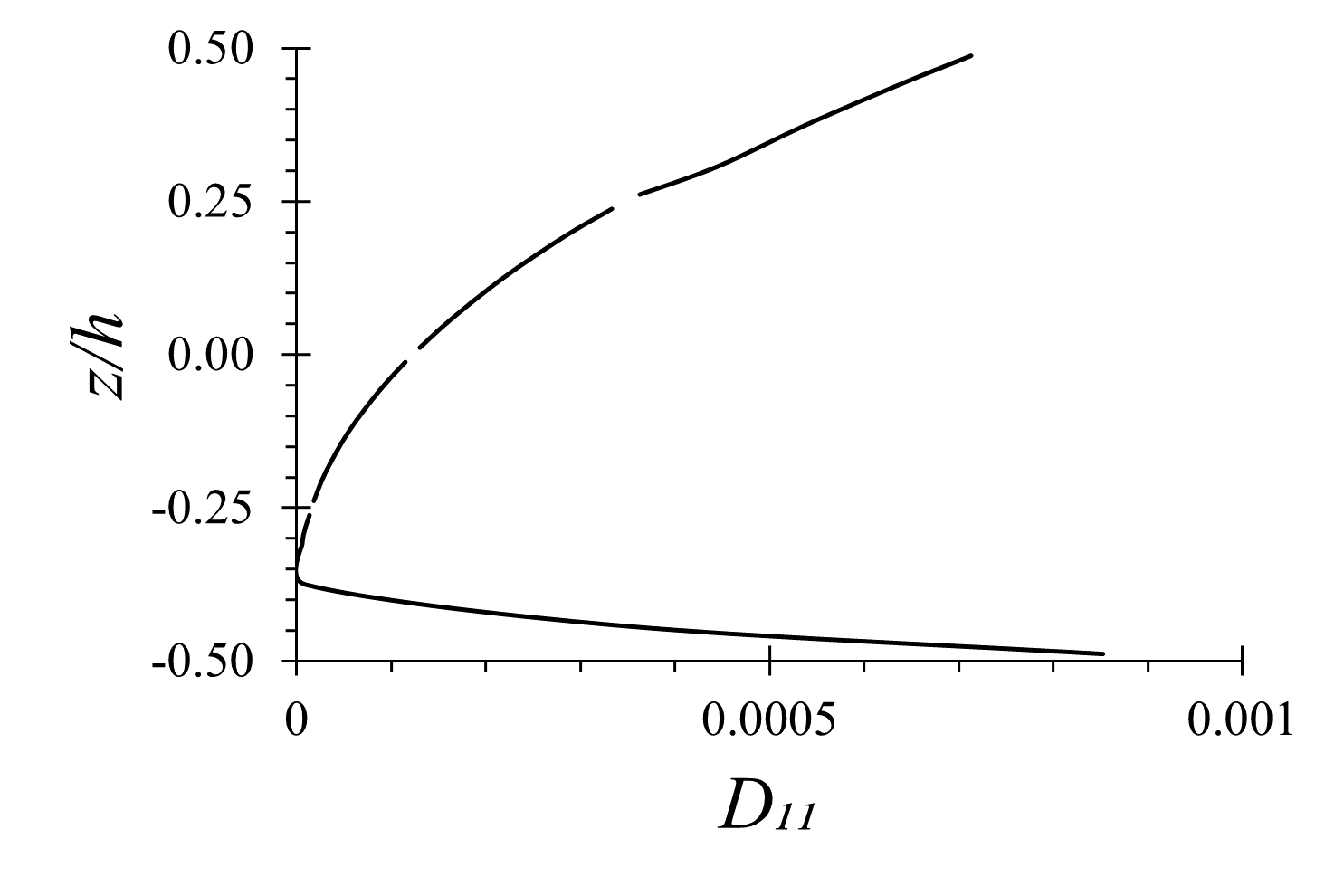}
\caption{}
\label{fig:d11c_mlttcm}
\end{subfigure}
\begin{subfigure}{0.5\textwidth}
\includegraphics[width=\textwidth]{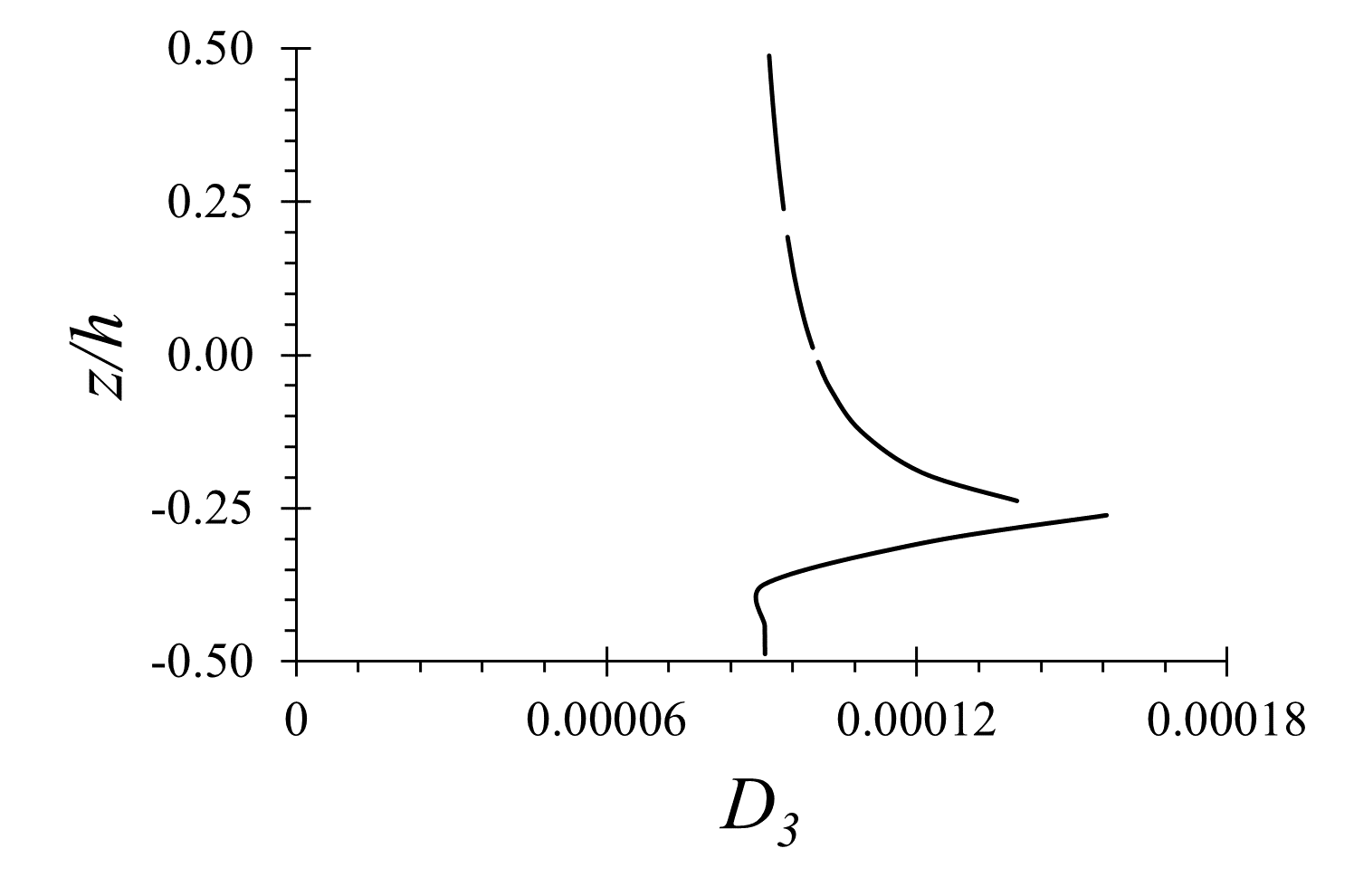}
\caption{}
\label{fig:d3c_barb}
\end{subfigure}
\begin{subfigure}{0.5\textwidth}
\includegraphics[width=\textwidth]{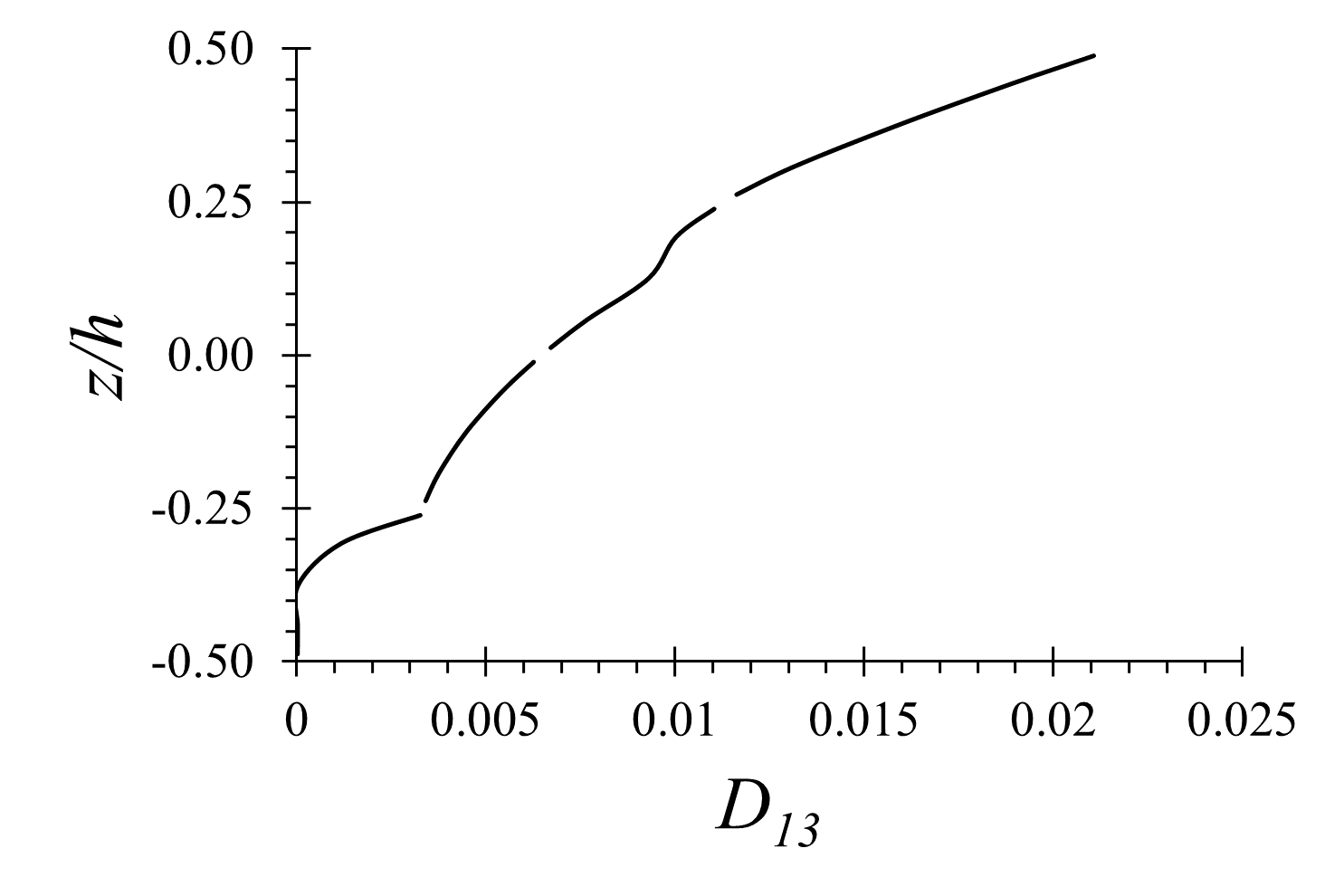}
\caption{}
\label{fig:d13c_mlttcm}
\end{subfigure}
\caption{{Through the thickness variation of damage variables $D_1,D_3$ and $D_{11},D_{13}$ at the center of the beam for (a),(c) Barbero and De Vivo} \cite{barbero2001constitutive} {associative damage model (Applied Load = 23.56 MPa) and (b),(d) Proposed damage model (Applied load = 24.39 MPa)} }
\label{fig:D vs Z}
\end{figure}

{The damage evolution equations of Barbero and De Vivo damage model} \cite{barbero2001constitutive} {and the present damage model involve coupling vectors which account for coupling among different damage variables at a material point. The coupling vector of Barbero and De Vivo associative damage model (Eqns. 9 and 14 of} \cite{robbins2005efficient}) {involve participation of in-plane normal strain $\varepsilon_{11}$ and out of plane shear strain $2\varepsilon_{13}$ for the evolution of damage variable $D_1$ and only out of plane shear strain for the evolution of damage variable $D_3$. This leads to more evolution of $D_1$ compared to $D_3$ as shown in} Fig.~\ref{fig:d1c_barb} and~\ref{fig:d3c_barb}. {However, the coupling vector of the proposed failure mode dependent associative damage model involves different constant terms (independent of stress and strain) with more evolution of $D_{13}$ compared to $D_{11}$ at a particular material point which is evident in Fig.}~\ref{fig:d11c_mlttcm} and~\ref{fig:d13c_mlttcm}. {However, the overall degradation of both the damage models is nearly same with Barbero and De Vivo model predicting slightly smaller stiffness degradation than the present model (Fig.}~\ref{fig:ldc}).

\begin{figure}[h!]
\begin{subfigure}{0.5\textwidth}
\includegraphics[width=\textwidth]{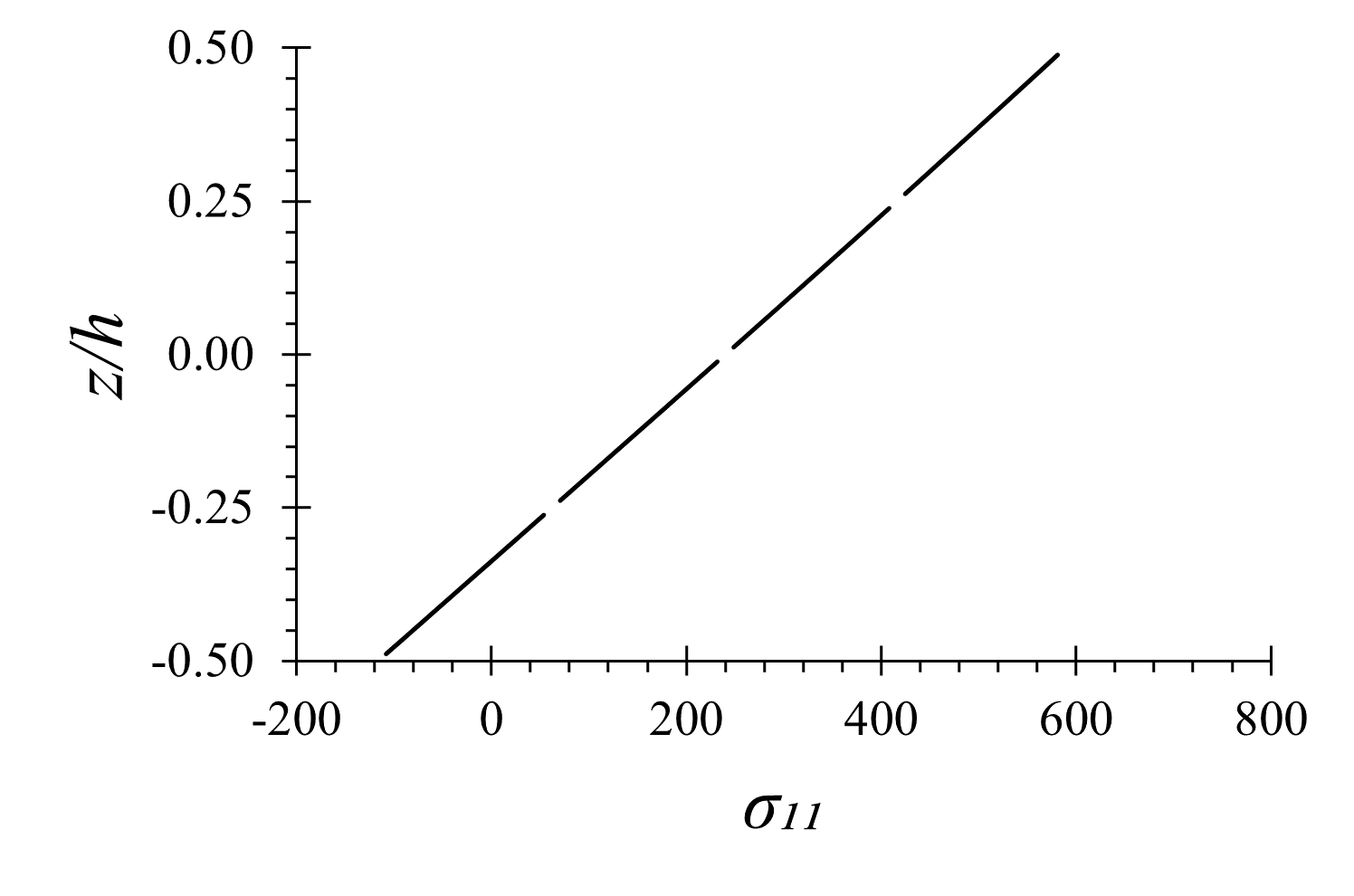}
\caption{}
\label{fig:s11c_barb}
\end{subfigure}
\begin{subfigure}{0.5\textwidth}
\includegraphics[width=\textwidth]{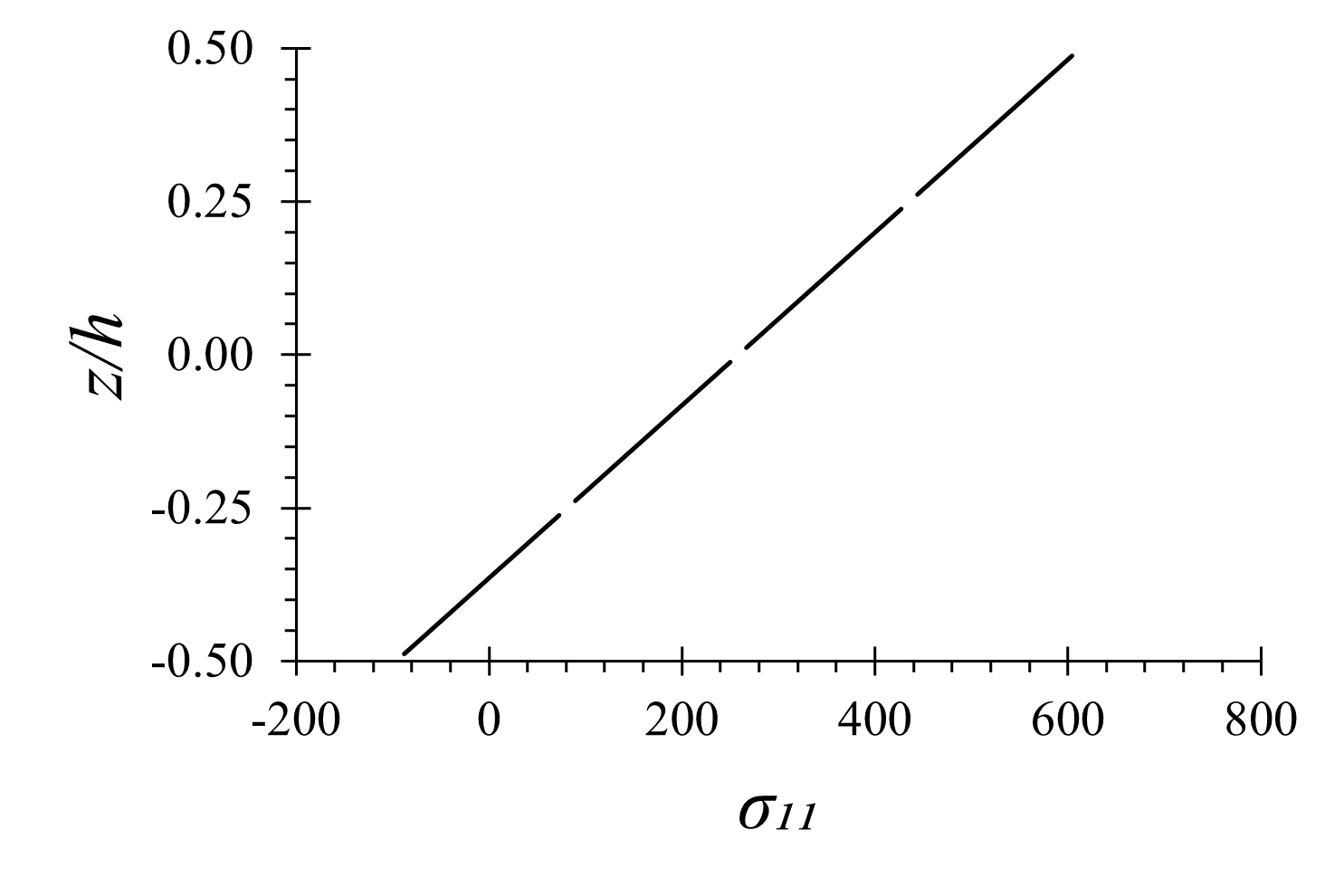}
\caption{}
\label{fig:s11c_mlttcm}
\end{subfigure}
\caption{ {Through the thickness variation of in-plane stress $\sigma_{11}$ vs $z/h$ at the center of the beam for (a) Barbero and De Vivo} \cite{barbero2001constitutive} {associative damage model (Applied load = 23.56 MPa) and (b) Proposed damage model (Applied Load = 24.39 MPa).} }
\label{fig:stress11 vs z}
\end{figure}

Through the thickness variation of the damage variable responsible for stiffness degradation in fiber direction (Fig.~\ref{fig:D vs Z}), at the center of the curved beam, depicts that the present damage model (Fig.~\ref{fig:d11c_mlttcm}) predicts the evolution of damage consistent with the strength properties of the material (Table~\ref{Material Properties}) with more damage in fiber direction compression compared to tension. However, Barbero and De Vivo's damage model is not able to capture the evolution of damage consistent with the strength properties of the material because the damage evolution equations are failure mode independent and driven by the stress magnitude ($\sigma_{11}$ more in fiber tension compared to fiber compression as shown in Fig.~\ref{fig:stress11 vs z}) whereas the proposed damage model takes into account the mode of failure in fiber reinforced composite materials.

\section{Conclusions}

A three-dimensional thermodynamically consistent and failure mode-dependent damage model with different damage hardening variables in each of the damage modes and polynomial-based damage hardening functions is proposed in this article for fiber-reinforced composite materials. The damage model parameters are characterized using the steepest descent optimization algorithm to minimize the difference between the model predicted and experimental uniaxial/shear stress-strain curves.
The optimized damage model is used for stiffness degradation prediction of a laminated composite curved beam subjected to a uniformly distributed transverse load. The proposed damage model is also compared with a failure mode independent Barbero and De Vivo \cite{barbero2001constitutive} associative damage model and {an existing failure mode dependent damage model} \cite{pan2022use}. Through the thickness variation of the damage variable at the center of the beam depicts that unlike Barbero and De Vivo damage model, the present damage model is able to capture evolution of damage consistent with the strength properties of the material depicting its failure mode dependent nature. {Also, the developed model predicts significant non-linearity in the load vs deflection curve, thereby accounting for stiffness degradation due to micro-crack evolution. The developed damage model can be coupled with cohesive zone modeling to account for delamination damage at the interface of the two laminae and will be taken up as a future work.}
%

%

 \bibliography{refrences}
%







\end{document}